# Unified Band Theoretic Description of Electronic and Magnetic Properties of Vanadium Dioxide Phases

Sheng Xu,[1,2] Xiao Shen,[2,3] Kent A. Hallman,[2] Richard F. Haglund, Jr.,[2]

Sokrates T. Pantelides[2,4]

[1]School of Metallurgy and Materials Engineering, Jiangsu University of Science and Technology,
Zhangjiagang, Jiangsu 215600, China
[2]Department of Physics and Astronomy, Vanderbilt University,
Nashville, TN 37235-1807, USA
[3]Department of Physics and Materials Science, University of Memphis,
Memphis, TN 38152, USA
[4] Department of Electrical Engineering and Computer Science, Vanderbilt University,
Nashville, TN 37235-1807, USA

The debate about whether the insulating phases of vanadium dioxide ($VO_2$) can be described by band theory or must be described by a theory of strong electron correlations remains unresolved even after decades of research. Energy-band calculations using hybrid exchange functionals or including self-energy corrections account for the insulating or metallic nature of different phases, but have not yet successfully accounted for the observed magnetic orderings. Strongly-correlated theories have had limited quantitative success. Here we report that, by using hard pseudopotentials and an optimized hybrid exchange functional, the energy gaps and magnetic orderings of both monoclinic $VO_2$ phases and the metallic nature of the high-temperature rutile phase are consistent with available experimental data, obviating an explicit role for strong correlations. We also report a potential candidate for the newly-found metallic monoclinic phase and present a detailed magnetic structure of the M2 monoclinic phase.

Vanadium dioxide ($VO_2$) exhibits a first-order phase transition from an insulating phase to a metallic phase at 340 K [1], which is accompanied by a structural transition from the monoclinic M1 phase to the tetragonal rutile (R) phase. $VO_2$ is intensively studied for such applications as temperature-tuned memory materials [2] and smart windows [4], and for optoelectronic devices [3] that take advantage of the ultrafast nature of this transition when excited electrically or optically. It is also widely viewed as a model system for understanding insulator-to-metal transitions in solids. The insulating M1 phase of $VO_2$ has an optical band gap of 0.6-0.7 eV [4,5] and can be considered nonmagnetic (NM) [6,7] near room temperature, while the metallic R phase is paramagnetic (PM) [4] above the transition temperature of 340 K. In addition to these two phases, the experimentally derived phase diagram of $VO_2$ [8,9] includes a second insulating monoclinic phase designated as M2, which can be stabilized in doped or strained $VO_2$ single crystals [10,11], thin films [12,13], and nanobeams [14]. Recently, stable metallic monoclinic (mM) phases were also found to exist near room temperature under high pressure [15] and in epitaxial thin films [16,17]. These phases may be related to the transient metallic monoclinic state already reported in ultrafast experiments [16,18]. This complexity makes a theoretical understanding of the $VO_2$ phases particularly interesting and important.

The theoretical description of $VO_2$ phases has been controversial for half a century. The debate has centered on the question whether the insulating phases can be described by single-quasiparticle band theory or the band gap results from strong correlations in the Mott-Hubbard sense [10,11,19,20]. In 1971, Goodenough suggested that the band gap in $VO_2$ can originate from the formation of V-V pairs [21], but, in 1975, Zylbersztejn and Mott proposed that the band gap in $VO_2$ originates largely from strong electron correlations [22]. This thesis subsequently gained support from experimental data that showed behavior similar to the generic, non-material-specific predictions of model correlated-electron Hamiltonians [19,23]. In 1994, density-functional theory (DFT) calculations for the M1 phase, based on the local density approximation (LDA) for the exchange-correlation potential, favored a Peierls-like dimerization of V atoms as the root of insulating behavior [24]. However, these DFT calculations did not yield a true band gap, a failure which strengthened arguments for a Mott-Hubbard description of the band gap [23,25]. In 2005, Biermann *et al.* carried out dynamical mean-field theory (DMFT) calculations, effectively building electron correlations into DFT-LDA calculations that give zero energy gap [26]. They found a nonzero band gap for the M1 phase, but concluded that $VO_2$ *is not a conventional Mott insulator*; instead the finite band gap was attributed to a *correlation-assisted Peierls transition*.

In the last decade, single-particle theories have been extensively explored and tested against experiment. In 2007, Gatti *et al.* [27] calculated $VO_2$ energy bands using Hedin's GW approximation for the one-electron Green's function [28], which replaces the bare Coulomb potential in the Hartree-Fock (HF) approximation by an energy-dependent screened Coulomb interaction. These calculations produced an energy gap in the M1 phase and a metallic rutile phase. In 2011, Eyert [29] reported energy-band calculations using then-newly-developed hybrid exchange-correlation functionals, in which a fraction of the local exchange potential is replaced by HF exchange. He obtained satisfactory energy gaps for the insulating phases, duplicating the success of Gatti *et al.* [27], and addressed the issue of magnetic ordering. While this initial success was followed by more comprehensive studies [30–32], at this juncture, no single exchange-correlation functional has been found that reproduces both the observed energy gaps and magnetic orderings of $VO_2$ phases, so that the applicability of band theory to $VO_2$ remains in dispute.

In this Letter, we introduce two novel elements in energy-band calculations for the principal phases of $VO_2$: (1) significantly harder pseudopotentials for both oxygen and vanadium and (2) an optimized mixing parameter in a hybrid functional for the exchange-correlation potential. The calculated lattice constants, band gaps, and magnetic properties of the R, M1 and M2 phases of VO$_2$ are consistent with available experimental data. Additionally, the calculated density of states (DOS) for the M1 phase is quantitatively consistent with experimental x-ray photoemission (XPS) data. The success of these hybrid DFT calculations demonstrates that band theory can describe VO$_2$ phases without explicitly invoking strong correlations. Moreover, the calculations predict a new monoclinic phase with a crystal structure that is intermediate between M1 and R, which we call the M0 state. The M0 phase is ferromagnetic and the true ground state of VO$_2$ at T = 0. Old data at liquid-helium temperature [33,34] suggest the existence of such a phase at near-zero temperatures, but more comprehensive data are needed to confirm the prediction. M0 may also be a candidate for the recently discovered [15–17] metallic monoclinic mM phase of VO$_2$ at finite temperatures.

Hybrid DFT calculations for each VO$_2$ phase were performed using a plane-wave basis and the projector-augmented-wave method [35] as implemented in the Vienna Ab initio Simulation Package (VASP) [36]. Several magnetic configurations were calculated to determine the magnetic ordering for each VO$_2$ phase. The exchange and correlation were described by a tuned PBE0 hybrid functional [37,38] that contains 7% HF exchange, which yields an energy gap for M1 in agreement with experiment. These calculations provide a more accurate description of the vanadium and oxygen atoms for two reasons. First, the oxygen pseudopotential in these calculations is harder than typically used (i.e., the core radius is smaller). As required by the harder oxygen pseudopotential, the plane wave cutoff energy is set at 700 eV; the use of 800 eV caused no appreciable changes. Secondly, thirteen electrons ($3s^23p^63d^44s^1$) were treated as valence electrons for vanadium instead of the typical eleven electrons [29,31]. For the oxygen atoms, six electrons ($2s^22p^4$) are treated as valence electrons as usual. Γ-centered k-point grids are used for all Brillouin zone sampling. We use a 3×3×3 grid for the M1 and M0 unit cells which each contain 12 atoms, a 4×4×6 grid for the R unit cell with 6 atoms, and a 1×2×2 grid for the M2 unit cell with 24 atoms. The self-consistent electronic calculations are converged to $10^{-4}$ eV between successive iterations, and the structural relaxations are converged so that the total-energy difference between two successive ionic steps is $10^{-3}$ eV.

The optimized crystal structures in Figure 1 agree well with the experimentally-derived structures [17,39–41]. All V-V chains of M1 and M0 are both canted and dimerized while R has undimerized straight V-V chains. The monoclinic M2 phase has both straight dimerized V-V chains and undimerized but canted antiferromagnetic V-V chains [11]. The crystal structure information gleaned from the calculations is listed in Table S1 [42].

First, we consider the magnetic and electronic properties of the R phase. Experiments have shown that the R phase is PM above the transition temperature of 340 K [4,43]. According to the present calculations, the total energies of antiferromagnetic R (AFM-R) and NM-R are higher than ferromagnetic R (FM-R) by ~125 and 140 meV per formula unit, respectively. Although the calculations predict FM-R to be the ground state of R, the temperature at which DFT calculation must be performed (0 K) is well below any hypothetical Curie temperature of R-VO$_2$. However, the crystal structure of VO$_2$ is monoclinic at temperatures below 340 K. Thus, we cannot directly compare the calculated FM ground state to an experimentally-observed state and can only state that our FM-R prediction is consistent with the experimental observations of PM-R [4,43]. As shown in Table 1, which compares band gaps and magnetic ground states of VO$_2$ phases, FM-R corresponds to a metal, in agreement with experiment [4,43] and a previous hybrid calculation [44], but unlike other hybrid calculations [31,32].

We next consider the magnetic and electronic properties of the M1 phase. Conflicting reports of paramagnetic [6] and diamagnetic [7] susceptibilities for M1 suggest that M1 probably has a negligible magnetic susceptibility, and that experimental values are potentially affected by fabrication parameters; we therefore designate it as NM as previous authors have done [31]. The optimized AFM-M1 spin configuration relaxes to the more stable NM-M1 in contrast to previous hybrid DFT results [29–31,44] but consistent with experiment [4,6,43]. On the other hand, optimizing the M1 crystal structure with initial ferromagnetic ordering relaxes to a new crystal structure as outlined in the next paragraph. As can be seen in Table 1, we obtain a band gap of 0.63 eV for NM-M1 in good agreement with the experimental value [4,5,45] of 0.6-0.7 eV and the values obtained from GW [27] and DMFT [26] calculations. In Figure 2, the total DOS of NM-M1 is compared to the experimental XPS spectra [45] and the GW DOS of Ref. [27]. The shape of the DOS and the positions of peaks from -10 to 0 eV agree well with the experimental results [45] and with the GW DOS. This comparison confirms that the electronic structure of the insulator phase NM-M1 is correctly reproduced by the present hybrid DFT calculations.

In addition to the NM-M1 and FM-R states, the present hybrid DFT calculations predict a stable ferromagnetic state, FM-M0, with a structure intermediate between NM-M1 and FM-R. Calculations starting from the FM-M1 configuration converge to FM-M0 during geometry optimization. Since the total energy of FM-M0 is lower than the calculated energy of the commonly accepted ground state, NM-M1, by ~50 meV per formula unit, we suggest that VO$_2$ is likely to be ferromagnetic at very low temperatures. A low Curie temperature could account for the discrepancy between the predicted ferromagnetism and the finite magnetic susceptibility observed in experiments at moderately low temperatures [33,34]. Between 30 K and the insulator-to-metal transition at ~340K the magnetic susceptibility is small [34], reinforcing the conventional wisdom that NM-M1 is the stable phase above 30 K. Calculations with initial configurations of AFM-M0 and NM-M0 both converge to NM-M1. Along with the fact that FM-M1 converges to FM-M0, these calculations hint at the complex interplay of magnetic and structural degrees of freedom, and highlight the necessity of more magnetic measurements at low temperatures to confirm previous experimental results [33,34] and test our theoretical predictions.

Similar to NM-M1, the FM-M0 configuration has a simple monoclinic lattice with space group P2$_1$/c (C$^5_{2h}$, No. 14) and dimerized zigzag V-V chains. However, the crystal structures of NM-M1 and FM-M0 exhibit subtle differences, as shown in Figures 1(a) and 1(b). The short V-V bond of FM-M0 is longer and the long bond is shorter than the corresponding bonds in NM-M1. Therefore, the FM-M0 crystal structure can be viewed as an intermediate state between the crystal structures of NM-M1 and FM-R. It is noteworthy that both the short and long V-V bonds of FM-M0 are closer to the bond length found in FM-R than their NM-M1 counterparts, indicating a FM-M0 intermediate state would be structurally closer to FM-R than to NM-M1. Furthermore, the 175° bond angle of FM-M0 is also closer to the 180° angle found in FM-R than the 166° angle of NM-M1. Diffraction measurements and optical or electrical measurements below the Curie temperature are needed to verify the structure and metallic character of the FM-M0 state.

Recently, a stable metallic monoclinic $VO_2$ phase (mM) has been observed near room temperature in epitaxial thin films [17] and single crystals under high pressure [15]. In the thin films, X-ray absorption fine-structure spectroscopy (XAFS) demonstrated that the short V-V bond elongates, the long V-V bond shortens, and zigzag V-V chains straighten when $VO_2$ metallizes [17], leading to an intermediate crystal structure with lattice constants and bond lengths nearly identical with those for FM-M0 shown in table S1 [42]. Pressure-dependent Raman spectroscopy, mid-infrared reflectivity, and optical conductivity measurements confirmed an insulator-to-metal transition without an accompanying structural transition from monoclinic to the rutile phase [15]. However, although a subtle change in structure was attributed to the appearance of the M2 phase, that assignment explains neither the metallization nor the fact that intermediate Raman spectra are unlike either M2 or M1 [15]. Instead, a monoclinic metallic phase, such as M0, with slightly different crystal structure than either M1 or M2, would explain both the mM phase in thin film samples [17] and the metallic monoclinic $VO_2$ phase that appears under high pressure [15]. The similar crystal structures and metallic character of the predicted FM-M0 and the experimental mM states suggest that FM-M0 may be related to this mM phase.

Although most work on $VO_2$ over the past fifty years has focused exclusively on the transition between the insulating M1 and metallic R phases, multiple authors [9,11,23,29,46] have suggested that the M2 insulating phase may hold the key to a complete understanding of the $VO_2$ phase transition. Three possible AFM configurations [47] designated as A-AFM, G-AFM, and C-AFM are shown in Figure 3(a), 3(b), and 3(c), respectively. Each configuration represents a unique magnetic ordering on the zigzag chains, while the straight chains have no moments. The A-type and G-type exhibit antiparallel moments along the canted zigzag V-V chains [11]. For A-AFM, moments on V-atoms in a canted zigzag chain are parallel to moments of its nearest V-atom neighbors on the next canted chain, while they are antiparallel for G-AFM and C-AFM. However, the moments of all vanadium atoms on a single chain are aligned in C-AFM.

Our calculations show that the A-AFM is the lowest energy configuration of M2 and the G-AFM, C-AFM, FM, and NM configurations of M2 are higher in energy than A-AFM by ~4 meV, ~27 meV, ~16 meV, and ~32 meV per formula unit, respectively. Although numerically accurate, the small energy difference (4 meV) between A-AFM and G-AFM may not be captured accurately by the approximate functionals. Nevertheless, both A-type and G-type AFM-M2 agree with the experimentally derived model in which M2 is antiferromagnetic and local magnetic moments are present only on the canted zigzag V-V chains [11]. Similarly, the present calculations show that the local magnetic moments of AFM configurations are on the canted V-V chains while the straight, dimerized chains have negligible moments. The band gap of 0.56 eV calculated for A-AFM-M2 is in agreement with photoelectron spectroscopy (PES) of M2 quoting a band gap greater than 0.1 eV [48]. Although PES is a surface-sensitive technique and vanadium dioxide films can be highly defective at the surface, our value of 0.56 eV confirms the band model proposed by Goodenough [49] in which the band gap for M2 is comparable to but smaller than the band gap of M1 (0.6-0.7 eV).

The kernel of the long-standing debate about $VO_2$ is whether the electronic properties of this material are better described by band theory in which electrons are represented by non-interacting quasiparticles that experience the same single-particle crystal potential, or by a many-body approach in which electron-electron interactions are explicitly incorporated.

In principle, band theory can always describe any given material: ground-state properties are describable by DFT, assuming that a satisfactory exchange-correlation potential $V_{xc}(r)$ can be constructed; excitations can be described by Hedin's GW expansion of the self-energy $\Sigma(r,r';E)$ followed by solving the Bethe-Salpeter equation [50] to include electron-hole interactions. Both the DFT and Hedin equations look like Schrödinger equations: the $V_{xc}(r)$ in DFT is replaced by the nonlocal, energy-dependent $\Sigma(r,r';E)$. In both cases, one gets quasiparticle energy bands, single-particle excitations and collective excitations (plasmons) from the zeros of the real part of the single-particle dielectric function [51], but the energy dependence in $\Sigma(r,r';E)$ is often essential [27]. The standard procedure is to first solve the DFT equation with a reasonable choice of $V_{xc}$, and then use the solutions to construct $\Sigma(E_k)$, which are in turn used to correct the DFT energy bands. Ideally, the process should be carried to self-consistency to eliminate the effect of the initial $V_{xc}$ choice. Gatti *et al.* [27] have already demonstrated that this process correctly predicts the band gap of insulating monoclinic $VO_2$, but the numerical procedures are quite cumbersome and magnetic calculations require separate, self-consistent GW calculations. Hybrid exchange-correlation functionals constitute an attempt to construct a $V_{xc}(r)$ that also serves as a local, energy-independent approximation to $\Sigma(r,r';E)$, known as the COHSEX approximation. The mixing parameter in the hybrid functional can be used to tune the functional to each material, as done in the present paper (a free parameter – the Hubbard U – is also present in theories that incorporate explicit electron-electron interactions). Here we have demonstrated that, by tuning the mixing parameter and using harder-than-usual pseudopotentials, the single-particle approach works to yield both the electronic and magnetic properties of $VO_2$ phases (the nature of the phase transition is not addressed here).

While DFT and GW serve as rigorous *quantitative* tests of quasiparticle theories, claims that $VO_2$ is a strongly-correlated material have been based primarily on model many-body Hamiltonians that are not material-specific. Experimental data in the region of the phase transition have been compared with the corresponding model behavior [20,23]. The appearance of correlated behavior at the phase transition, however, does not necessarily imply that strong correlations persist at temperatures away from the phase transition. Quantitative theories based on strong correlations, such as LDA+U, GGA+U and DMFT, assume at the outset that strong electron-electron interactions, incorporated via the Hubbard-model on-site parameter U, dominate. In the case of $VO_2$, LDA+U yields insulating behavior for both the monoclinic and rutile phases [52,53]. The DMFT calculations by Biermann *et al.* are anchored on a zero-gap DFT calculation and found that the Hubbard U is needed to get the observed value of a Peierls-induced energy gap. These calculations, however, attributed a low-energy feature in the PES data to a lower Hubbard band, whereas the GW calculations by Gatti *et al.* [27] showed that the observed feature is actually a plasmon. It is clear that a DMFT calculation anchored on the present hybrid-functional DFT calculation, which already produces the observed energy gap, would have to use a near-zero U to reproduce experimental data. The results of Ref. [27] also suggest that the hybrid functional used here captures roughly the same physics as the GW/COHSEX calculation, whereby a GW calculation (at $G_0W_0$ level) on top of the present results would keep the same energy gap and improve the excitation spectra.

In summary, we have successfully reproduced the electronic and magnetic properties of M1, M2, and R phases of $VO_2$ using DFT calculations with a hybrid functional and accurate pseudopotentials. The success of these hybrid DFT calculations suggests that band theory can provide an adequate description of $VO_2$ phases. Moreover, the present calculations predict a new monoclinic ferromagnetic metal state of $VO_2$, which accounts for the magnetic data at low temperature and is also a candidate for the recently observed metallic monoclinic mM phase that appears in thin films or under high pressure. In addition, the antiferromagnetic structure of M2 was determined to be A-type. Verification of ferromagnetism in room-temperature $VO_2$ under high pressure as well as structural and electronic measurements at low-temperature in unstrained $VO_2$ are important future experimental directions to confirm the validity of our findings. In conclusion, our study underlines the power of the hybrid DFT approach to produce a comprehensive theoretical picture of *all* the $VO_2$ equilibrium phases and their magnetic properties.

**Acknowledgments:**

Research at Vanderbilt was supported by National Science Foundation grants DMR-1207241 and EECS-1509740 (KAH), by Department of Energy grant DE-FG02-09ER46554, and by the McMinn Endowment at Vanderbilt University. Computational resources were provided by the NSF XSEDE under grants TG-DMR130121, TG-DMR150028, and TG-DMR150063, and by the High Performance Computing Facilities at the University of Memphis. S. Xu was supported by the Jiangsu Overseas Research & Training Program for University Prominent Young & Middle-Aged Teachers and Presidents (China). We thank Lucia Reining for valuable comments on the manuscript.

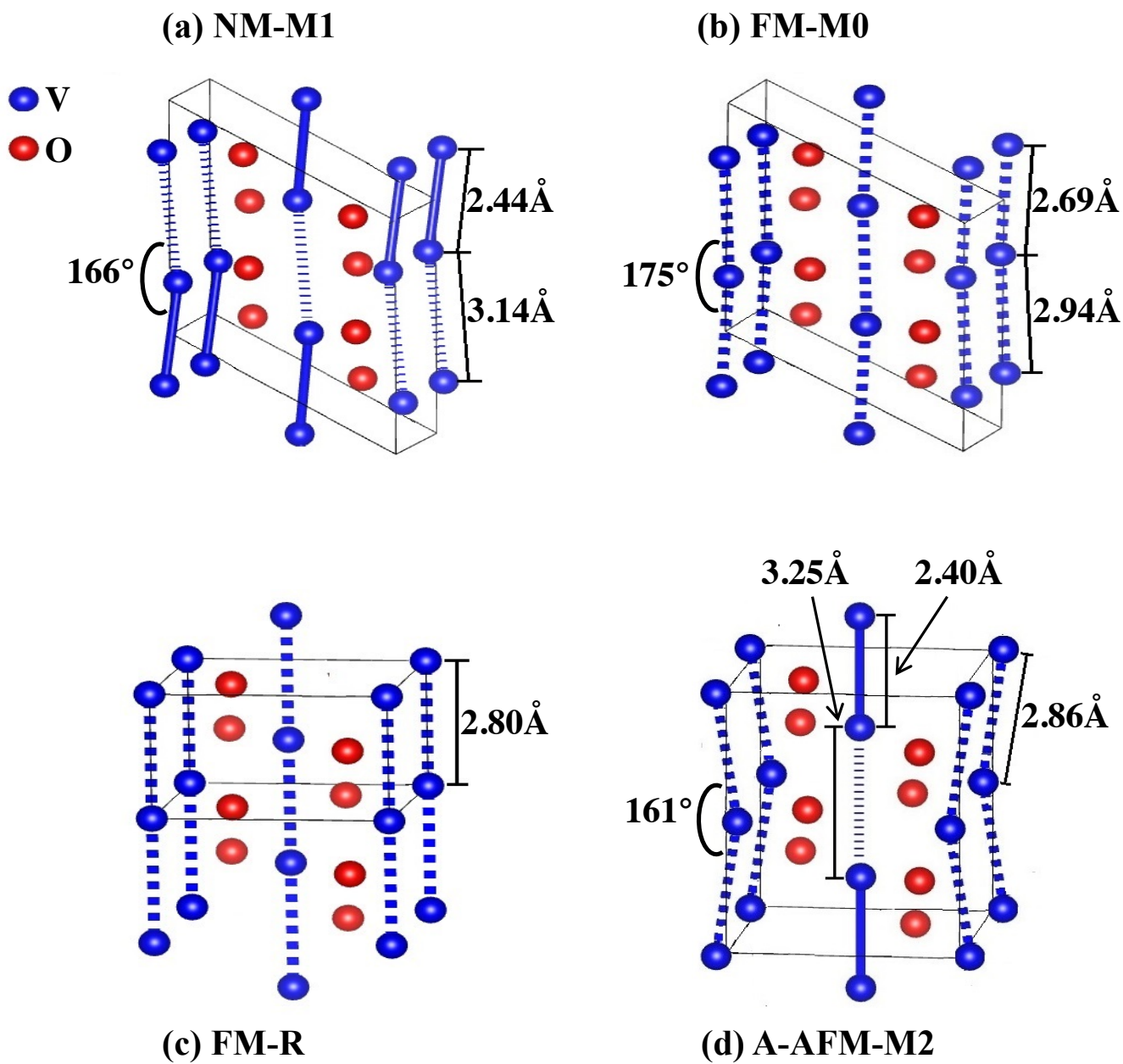

**FIG. 1.** Optimized structures of different VO$_2$ phases: (a) NM-M1, (b) FM-M0, (c) FM-R, and (d) A-AFM-M2. Short V-V bonds (<2.50Å) are shown as solid lines ( ) while long bonds (>3.00Å) have dotted lines ( ). V-V bonds with lengths between 2.50 and 3.00Å have dashed lines ( ).

**Table 1.** Calculated magnetic grounds states and band gaps of VO$_2$ phases compared to experiment.

| | | Experiment | Theoretical results | | | | | |
|---|---|---|---|---|---|---|---|---|
| | | | This work | HSE | | | GW | DMFT |
| | | | | [29][c] | [30][d] | [29] | [27] | [26][g] |
| Magnetic ground states | M0 | FM/PM [33,34][a] | FM | | | | | |
| | M1 | NM [6,7][b] | NM | | AFM | AFM | | |
| | M2 | AFM [11] | A-AFM | | | FM | | |
| Band gap (eV) | M1 | 0.6-0.7 [4,5] | 0.63 | 1.10 | 2.23 (AFM) 0.98 (NM)[e] | | 0.65 | 0.60 |
| | M2 | >0.10 [48] | 0.56 | 1.20 | | | | |
| | R | 0 [4,5] | 0 | 0 | 1.43 (FM) 0 (NM)[f] | | 0 | 0 |

(a) Divergence of the magnetic susceptibility below 30 K underlines the importance of exploring the unknown low-temperature magnetic properties.
(b) The disagreement of measurements of small positive [7] susceptibility and another publication [8] reporting small negative susceptibility justified our designation of M1 as NM as similar to previous authors [30].
(c) Band gap of each VO$_2$ phase was calculated by assuming the magnetic state found in experiments.
(d) Non-spin-polarized calculations similar to those of Eyert [29] were reproduced and then spin-polarized calculations for each potential magnetic state were performed [30].
(e) The correct magnetic phase, NM-M1, has a calculated band gap is close to the experimental value. However, AFM-M1 was calculated to be lower in energy, and the band gap is over thrice the expected value.
(f) A ferromagnetic R state with a band gap of 1.43 eV was calculated to be the ground state. However, a NM

— Page 11 —

state with a correct band gap of 0 was also obtained, albeit at a higher energy.

(g) A stable nonmagnetic structure was obtained with cluster-DMFT, but it was not compared to other magnetic states to determine the ground state.

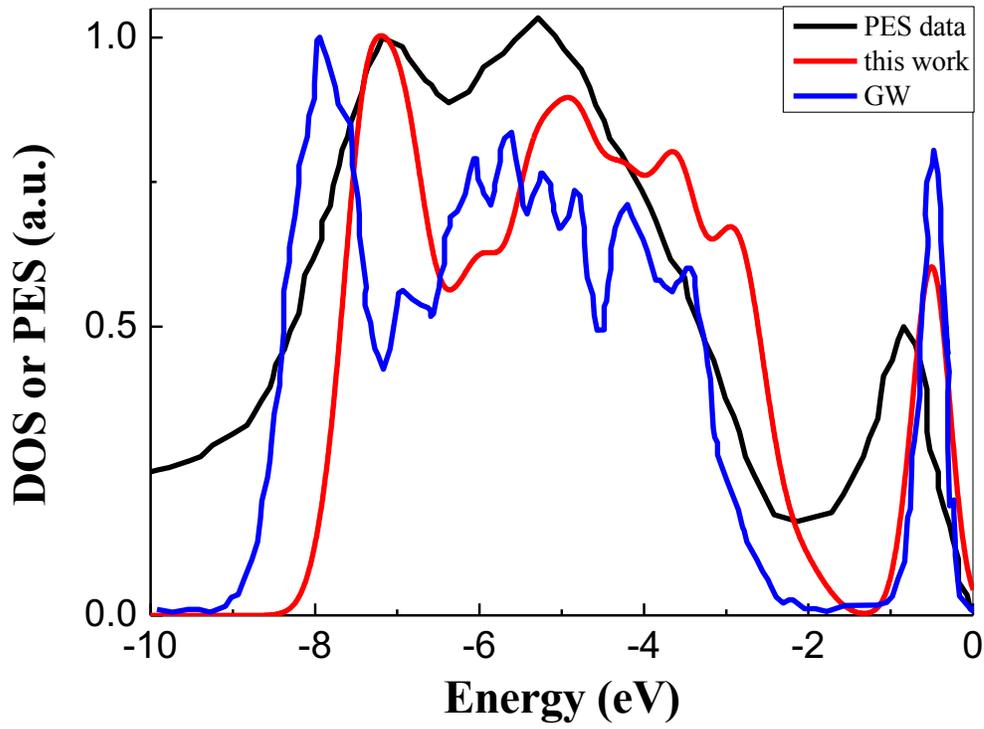

**FIG. 2.** The total DOS of NM-M1 calculated in this work (red) is compared with the experimental [45] photoemission spectrum (black) of the low temperature insulating M1 and the DOS (blue) from GW calculations [26]. The DOS from this work was convoluted with a Gaussian function.

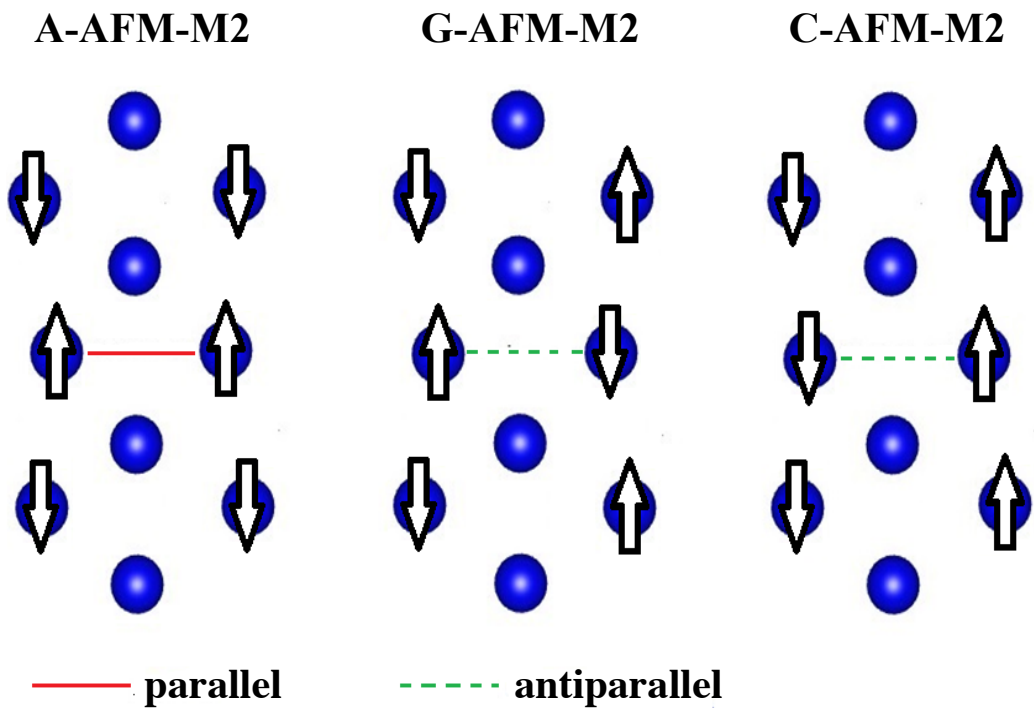

**FIG. 3.** Schematic of the 3 potential magnetic structures of AFM-M2: A-AFM, G-AFM and C-AFM. The blue solid circles are V atoms and the white arrows represent their moments. The solid line between two adjacent canted chains represents parallel magnetic moments between the nearest vanadium atoms from each chain, while the dashed lines represent an antiparallel configuration. A-AFM is the lowest in energy.